
\magnification=1200
\baselineskip=24pt plus 2pt
\centerline{\bf Comment on The Preprint}
\centerline{\bf ``Neutrino Flavor Evolution Near A Supernova's Core''}
\centerline{\bf (Astro-Ph 9405008) by J. Pantaleone}
\centerline{Yong-Zhong Qian and George M.
Fuller\footnote{$^1$}{Permanent address: Department of Physics,
University of California, San Diego, La Jolla, CA 92093-0319.}}
\centerline{Institute for Nuclear Theory}
\centerline{University of Washington, Seattle, WA 98195}
\centerline{Abstract}
The revised version of the widely circulated preprint ``Neutrino Flavor
Evolution Near A Supernova's Core'' by J. Pantaleone (astro-ph 9405008
on the Bulletin Board, Indiana University preprint 
IUHET-276) is wrong. It contains two errors which lead to incorrect
conclusions regarding neutrino flavor transformation in the supernova
environment. In this short note we discuss these errors.
\vfill
\eject
The widely circulated preprint ``Neutrino Flavor Evolution Near A
Supernova's Core'' by J. Pantaleone [1] is wrong. We pointed out that
the first version of this paper had a quantum mechanics error. In
response the author has changed the entire second half of the paper.
Unfortunately, this revised version still contains two errors which
render its conclusions incorrect.

First, Pantaleone makes a conceptual error in his treatment of the
neutrino density matrix. Individual neutrinos emitted from the neutrino
sphere can be described as coherent states (kets). However, each
emitted neutrino is related to every other emitted neutrino in an
incoherent fashion. These states have random relative phases, as is
characteristic of a thermal emission process. The total neutrino field
is properly a {\bf mixed} ensemble, {\bf not} a coherent many-body
state. The total neutrino density matrix is an incoherent sum over each
{\bf single} neutrino density matrix. The single neutrino density
matrix can be written as
$$\eqalignno{|\psi(t)\rangle\langle\psi(t)|&=|a_1(t)|^2
|\nu_1(t)\rangle\langle\nu_1(t)|+|a_2(t)|^2
|\nu_2(t)\rangle\langle\nu_2(t)|\cr
&+a_1^*(t)a_2(t)|\nu_2(t)\rangle\langle\nu_1(t)|+a_1(t)a_2^*(t)|\nu_1(t)
\rangle\langle\nu_2(t)|,
&(1)\cr}$$
where the single neutrino state is
$|\psi(t)\rangle=a_1(t)|\nu_1(t)\rangle+a_2(t)|\nu_2(t)\rangle$, and
where $|\nu_1(t)\rangle$ and $|\nu_2(t)\rangle$ are the propagating
physical mass eigenstates for the case of two-neutrino mixing. Here $t$
represents any evolutionary parameter (e.g., density, radius, time,
etc.) along the neutrino's path. Note that the last two terms in Eq.
(1) have coefficients $a_1^*(t)a_2(t)$ and $a_1(t)a_2^*(t)$. These are cross terms.

These cross terms contain phases which cause them to be rapidly varying
with position above the neutrino sphere. Each cross term is
proportional to a factor $\sim\exp[i\int\omega_{12}(t)dt]$, with
$\omega_{12}$ the difference in the neutrino-flavor-oscillation
frequencies of the two mass eigenstates $|\nu_1(t)\rangle$ and
$|\nu_2(t)\rangle$. These oscillation frequencies are, in turn,
dependent on density. Near the neutrino sphere it can be shown that
$\omega_{12}\sim\sqrt{2}G_{\rm F}n_e$, with $n_e=Y_e\rho N_{\rm A}$ the
net electron number density [2, 3]. 

At a point above the neutrino sphere one must average over the neutrino
distribution functions in order to obtain an ensemble average for any
physical quantity dependent on the local neutrino configurations. In
any such ensemble average we will have to sum the individual neutrino
contributions (Eq. [1]) over different neutrino paths from the neutrino
sphere. Fig. 1 illustrates the arrangement of the neutrino sphere
(radius $R_\nu$) and a point at radius $r$ above it. Three possible
neutrino paths to the point at radius $r$ are shown. Each path with a
different polar angle will have a different phase entering into the
cross term coefficients of Eq. (1). The phase difference acquired from
going through a region of density $\rho$ with a path length difference
$\delta r$ is $\delta\phi\sim\sqrt{2}G_{\rm F}Y_e\rho N_{\rm A}\delta
r\sim 19(Y_e/0.5)(\rho/10^{10}\ {\rm g\ cm^{-3}})(\delta r/1\ {\rm
cm})\gg 1$. Clearly, the cross terms in Eq. (1) will average to zero in
any ensemble average over neutrino distribution functions.

Pantaleone mistakenly retains these cross terms in his expressions for
the ensemble-averaged neutrino density matrix elements (his Eqs. [13]
and [14]). This introduces a spurious, and unphysical, ``coherence''
which leads Pantaleone to the erroneous conclusion that the cross terms
dominate the neutrino flavor transformation phenomenon for cases
intermediate between the adiabatic and non-adiabatic limits.

In a second error, Pantaleone incorrectly estimates the effect of the
neutrino background on the adiabaticity of neutrino flavor evolution at
resonance. The full flavor-basis Hamiltonian for a neutrino propagating
above the neutrino sphere is (cf. Ref. [3]),
$$H={1\over2}
\pmatrix{-\Delta\cos2\theta+A+B&\Delta\sin2\theta+B_{e\mu}\cr
\Delta\sin2\theta+B_{e\mu}&\Delta\cos2\theta-A-B\cr},\eqno(2)$$
where $\Delta\equiv\delta m^2/2E_\nu$ with $\delta m^2=m_2^2-m_1^2$ the
difference of the squares of the vacuum mass eigenvalues and $E_\nu$ is
the neutrino energy. Note that Pantaleone's $\Delta$ [1] is our $\delta
m^2$. Here $\theta$ is the vacuum mixing angle in the unitary
transformation between mass eigenstates and flavor eigenstates in
vacuum. In this expression the contribution to the Hamiltonian from the
net electron number density is $A=\sqrt{2}G_{\rm F}n_e$. The
contributions to the Hamiltonian from neutrino-neutrino forward
scattering on background neutrinos are $B$ and $B_{e\mu}$ (Eqs. [15a \&
b] in Ref. [3]). With these definitions the adiabaticity parameter at resonance is
$$\gamma={(\Delta\sin2\theta+B_{e\mu})^2\over\Delta\cos2\theta}\left|
{d\ln(A+B)\over dr}\right|^{-1}_{\rm res}.\eqno(3)$$
Resonance occurs when $\Delta\cos2\theta=A+B$.

We are interested in estimating $\gamma$ at resonance for a given, {\it
fixed}, value of $\Delta$ and a given, {\it fixed}, value of
$\sin2\theta$. The term $B$ has the effect of shifting the resonance
position. The terms $B, \ B_{e\mu}$, and $|d\ln(A+B)/dr|^{-1}_{\rm
res}$ are to be evaluated at resonance. The essential nonlinearity of
this problem demands that a self-consistent iteration be performed to
obtain the resonance position and good estimates of $B$ and $B_{e\mu}$.

Instead of doing this, Pantaleone mistakenly takes $B$ to produce a
change in $\Delta$. By doing this he estimates a $\gamma$ which is {\it
irrelevant} for the resonance position of a neutrino with the original
values of parameters $\Delta$ and $\sin2\theta$. This is evident in his
Eq. (23) [1]. With these unphysical estimates for $\gamma$ Pantaleone's
conclusions regarding the flavor conversion efficiencies for neutrinos
with given $\Delta$ and $\sin2\theta$ are wrong.

Furthermore, Pantaleone incorrectly interprets Fig. 2 of Ref. [2]. He
states in Ref. [1] that ``... when the neutrino background is neglected
the flavor evolution is extremely adiabatic for almost all of the
relevant parameter space, and also because neutrino masses are only
probed if significant amounts of flavor conversion occur.'' This is
obviously false, as is revealed by a cursory estimate of flavor
conversion efficiency along the $Y_e=0.5$ line in Fig. 2 of Ref. [2]
(it is only of order $\sim 30$\% for $E_\nu=25$ MeV neutrinos).

The conclusions of Pantaleone's preprint regarding the effects of the
neutrino background on neutrino flavor transformation are in error. His
conclusion that ``For $\Delta<700$ eV$^2$ the connection between
$r$-process nucleosynthesis and cosmologically relevant neutrino masses
is enervated'' is wrong. A consistent treatment of neutrino flavor
evolution in supernovae, along with a proper ensemble average over
neutrino background quantities is performed in Ref. [3]. There it is
shown that the neutrino background produces only minor alterations in
the results of Ref. [2].

We would like to acknowledge discussions with A. B. Balantekin and W.
C. Haxton. This work was supported by the Department of Energy under
Grant No. DE-FG06-90ER40561 at the Institute for Nuclear Theory and by
NSF Grant No. PHY-9121623 at UCSD.
\vfill
\eject
\centerline{\bf References}
\vskip .2in
\noindent [1] J. Pantaleone, Indiana University Preprint No. IUHET-276
(1994) (astro-ph 9405008 on the bulletin board).

\noindent [2] Y.-Z. Qian, G. M. Fuller, G. J. Mathews, R. W. Mayle, J.
R. Wilson, and S. E. Woosley, Phys. Rev. Lett. {\bf 71}, 1965 (1993).

\noindent [3] Y.-Z. Qian and G. M. Fuller, Institute \ for \ Nuclear \
Theory \ Preprint \ No. DOE/ER/ 40561-150-INT94-00-63 (1994).
\vskip 1in
\centerline{\bf Figure Caption}
\vskip .2in
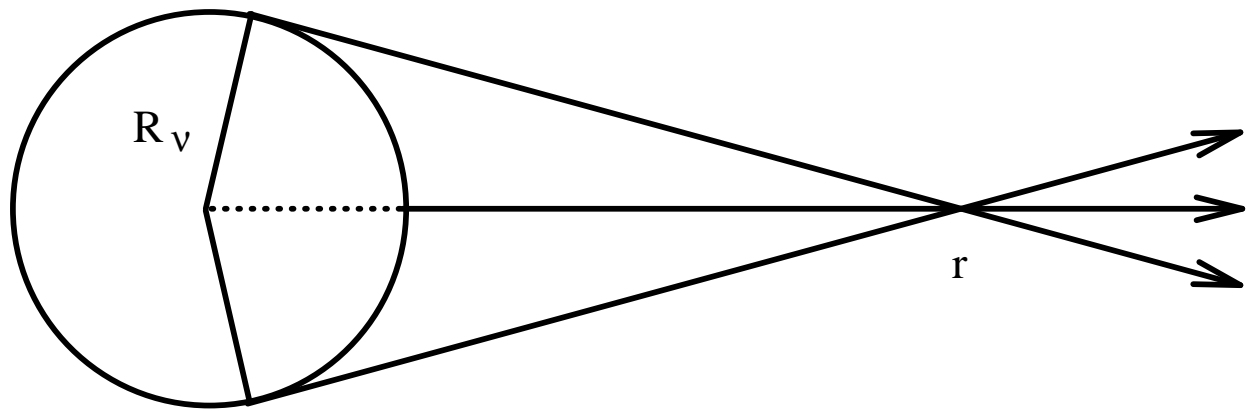
\noindent{\bf Fig. 1} Illustration of the arrangement of the neutrino
sphere (radius $R_\nu$) and a point at radius $r$ above it. Three
possible neutrino paths to the point at radius $r$ are shown.
\vfill
\eject
\end